\documentclass[aps,prapplied,reprint,superscriptaddress,floatfix,nofootinbib,showkeys]{revtex4-2}
\usepackage[utf8]{inputenc}
\usepackage{amsfonts,amssymb,amsmath}
\usepackage[hidelinks]{hyperref}
\usepackage{graphicx}
\usepackage{color,soul}

\newcommand{\mr}[1]{\mathrm{#1}}
\newcommand{\mb}[1]{\mathbf{#1}}
\newcommand{\be}{\begin{equation}}
\newcommand{\ee}{\end{equation}}

\newcommand{\rt}{R_\mr{T}}

\newcommand{\vob}{V_\mathrm{0b}}
\newcommand{\vb}{V_\mathrm{b}}
\newcommand{\ab}{A_\mathrm{b}}
\newcommand{\nng}{n_\mathrm{g}}
\newcommand{\nog}{n_\mathrm{0g}}
\newcommand{\vg}{V_\mathrm{g}}

\newcommand{\ag}{A_\mathrm{g}}
\newcommand{\cg}{C_\mathrm{g}}
\newcommand{\uv}{\,\mr{\mu V}}
\newcommand{\MHz}{\,\mr{MHz}}

\newcommand{\ec}{E_\mathrm{c}}

\newcommand{\LR}{\mr{L/R}}

\begin{document}

\title{Suppression of back-tunnelling events in hybrid single-electron turnstiles by source-drain bias modulation}
\author{Marco Marín-Suárez}\email{marco.marinsuarez@aalto.fi}
\affiliation{Pico group, QTF Centre of Excellence, Department of Applied Physics, Aalto University, FI-000 76 Aalto, Finland}
\author{Yuri A. Pashkin}
\affiliation{Department of Physics, Lancaster University, Lancaster LA1 4YB, UK}
\author{Joonas T. Peltonen}
\author{Jukka P. Pekola}
\affiliation{Pico group, QTF Centre of Excellence, Department of Applied Physics, Aalto University, FI-000 76 Aalto, Finland}

\begin{abstract}
Accuracy of single-electron currents produced in hybrid turnstiles at high operation frequencies is, among other errors, limited by electrons tunnelling in the wrong direction.
Increasing the barrier transparency between the island and the leads, and the source-drain bias helps to suppress these events in a larger frequency range, although they lead to some additional errors.
We experimentally demonstrate a driving scheme that suppresses tunnelling in the wrong direction hence extending the range of frequencies for generating accurate single-electron currents.
The main feature of this approach is an additional AC signal applied to the bias with twice the frequency as the one applied to the gate electrode.
This allows additional modulation of the island chemical potential.
By using the new approach under certain parameters, we improve the single-electron current accuracy by one order of magnitude.
Finally, we show through experimentally-contrasted calculations that our method can improve accuracy even in devices for which the usual gate driving gives errors $\sim 10^{-3}$ at high frequencies and can bring them under $5\times 10^{-4}$.
\end{abstract}

\maketitle

\section{Introduction}
Since the revision of the SI (\textit{système international d'unités})~\cite{CGPM2019}, quantum devices have been used for realizations of base units in terms of fundamental physical constants, especially in the electrical domain~\cite{Krasnopolin2002,Hohls2012,RibeiroPalau2015,Bae2020}.
For current standard implementations, mainly single-electron transport experiments have been carried out~\cite{Pekola2013,Kaestner2015,Yamahata2016}.
In particular, SINIS (S stands for superconductor, I for insulator, N for normal metal) single-electron turnstiles (SET) have been widely implemented as sources of stable DC currents~\cite{Pekola2007} among other applications~\cite{Kafanov2009,Knowles2012,Marin-Suarez2020} including a power standard~\cite{Marin-Suarez2022}.
In order to provide a reliable standard for the ampere, accuracy in generating a current $I=ef$, with $e$ the elementary charge and $f$ the turnstile operation frequency, besides high magnitude (at least $100\,\mr{pA}$) are needed.
Deviations from the desired current arise from tunnelling errors such as missed tunnelling events~\cite{Peltonen2015}, two-electron Andreev reflection~\cite{Aref2011,Maisi2014}, inelastic co-tunnelling~\cite{Averin2008}, quasiparticle poisoning~\cite{Knowles2012,Marin-Suarez2020,Taupin2016,Khaymovich2016,Zanten2016} and junction sub-gap leakage~\cite{Lotkhov2009,Pekola2010,Saira2010,Kemppinen2011} among other photon-assisted processes due to the effect of the electromagnetic environment~\cite{Marco2015}.

In the past, many proposals to minimize these errors have been put forward, mainly through device engineering.
Superconducting quasiparticle poisoning has been reduced by changing the lead geometry and superconducting energy gap engineering~\cite{Yamamoto2006,Knowles2012,Peltonen2017}, among others~\cite{Taupin2016,Marin-Suarez2020}.
Efforts for correcting missed tunnelling events have been directed towards error counting~\cite{Peltonen2015,Wulf2013,Fricke2014,Reifert2021}.
Andreev tunnelling can be suppressed by increasing the island charging energy $\ec$ so that it is larger than the energy gap $\Delta$ of the leads, as well as by increasing the tunnel barrier resistance~\cite{Kemppinen2009a,Aref2011}.
Furthermore, proper engineering of the electromagnetic environment has proved to reduce the junction sub-gap leakage~\cite{Lotkhov2011,Bubanja2014} and was proposed for suppressing inelastic co-tunnelling~\cite{Bubanja2011}.
One additional important error is the tunnelling of electrons in the direction opposite to the bias voltage~\cite{Marin-Suarez2022}.
These undesired events give $I<ef$ and appear when the amplitude of the driving signal is large enough and its rate of change (proportional to $f$) is comparable to the device response scale.
Typical approaches to avoid these include the decrease of either the total tunnel resistance of the device $\rt$ or $\ec$, however, this may result in an enhancement of two-electron Andreev events.
Furthermore, increasing the DC source-drain bias voltage also helps to suppress these events, but at the same time promotes sub-gap leakage current.
As an alternative, in the present work we demonstrate and justify a new driving method that extends the useful driving frequency range of a SINIS SET by suppressing electron tunnelling in the wrong direction and favors desired events.
Our approach mainly consists of adding a periodic modulation to the source-drain bias synchronized with the gate voltage modulation.

Our method effectively modulates the rate at which the island chemical potential crosses the energy thresholds that trigger tunnelling events.
This allows to increase the time window for tunnelling, favoring the desired events and effectively blocking the unwanted ones.
Particularly, we slow down the evolution of energy difference for the wanted processes while accelerating that of the unwanted events, hence decreasing the likelihood of the latter ones.
Recently, a similar approach has been employed to suppress back-tunnelling events in semiconductor quantum dot single-electron pumps~\cite{Hohls2022}, but with no bias modulation.
With this, we go beyond the device and setup optimization for error suppression, turning instead our attention to modifying the island chemical potential evolution beyond the simple gate waveform modification.

This article is divided as follows.
In Section \ref{Sec:Methods} we present the experimental methods used for the production of this work results.
The DC and AC characterization of the device under test is shown in Section \ref{Sec:Char}.
Next, in Section~\ref{Sec:Effect} we present the physical reasoning behind the results of this paper.
Furthermore, the results of this work are presented in Section~\ref{Sec:Results}.
Finally, Section~\ref{Sec:Conclusions} presents the conclusions of this article.

\section{Experimental methods}\label{Sec:Methods}
\subsection{Fabrication}
\begin{center}
\begin{figure}
\includegraphics[scale=0.8]{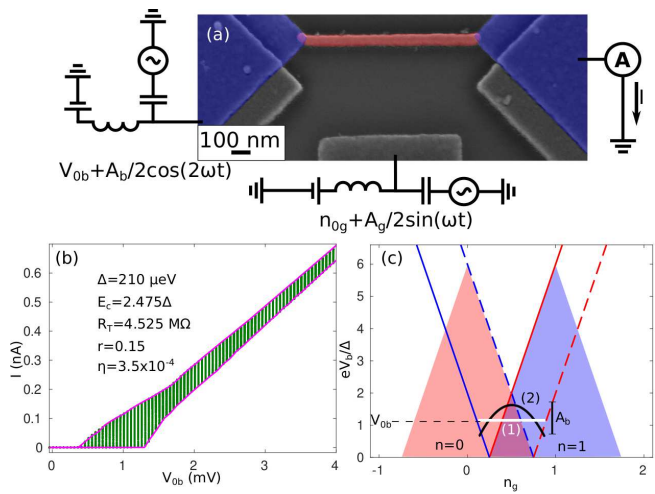}
\caption{(a) False-color electron micrograph with the measurement circuit used.
Red is normal-metal, blue is superconductor.
$\vob$, DC bias voltage; $\ab$, peak-to-peak (pp) bias amplitude; $\nog$, normalized DC gate-induced island charge; $\ag$, normalized pp gate charge amplitude; $I$, generated current; $A$, ammeter.
(b) DC IV curve.
Green vertical lines are measured data within several gate periods and magenta lines are maximum and minimum current calculated with a Markovian model.
(c) Top half of the stability diagram for a SINIS SET with $r=1$.
Inside the Coulomb diamonds, ideally, no current flows in the DC regime.
Light red diamond corresponds to the island with excess charge $n=0$ and light blue diamond to $n=1$.
Lines are tunnelling thresholds, blue (red) corresponds to right (left) junction, continuous (dashed) lines correspond to tunnelling into (out of) the island.
Crossing the line adds or removes one electron to or from the island.
(1) is the flat driving, (2) designates the parabolic driving.}
\label{f1}
\end{figure}
\end{center}
In order to put into practice our new method we fabricated the device of Fig.~\ref{f1}(a).
The normal metal island (red in Fig.~\ref{f1}(a)) is made of copper and the superconducting leads (blue in Fig.~\ref{f1}(a)) are made of aluminum.
We use e-beam lithography (EBL, Vistec EBPG5000+ operating at 100 kV) for sample patterning and e-beam metal evaporation for the metallization of the devices.
We do the only EBL step on a Ge-based hard mask~\cite{Meschke2016} deposited on top of a silicon-oxide coated silicon wafer.
This mask is composed of a layer of $400\,\mr{nm}$ of poly(methyl methacrylate-methacrylic acid) (P(MMA-MAA)) covered by a $22\,\mr{nm}$ layer of Ge which is deposited by e-beam evaporation, there is a final top layer of poly(methyl methacrylate) (PMMA).
In this patterning step, the small island and contacts of the SINIS transistor are drawn together with large bonding pads and connections between these and the small features.
For easier handling, the full 4 inch wafer is cut into smaller chips containing several devices.
After development of the top PMMA layer, the pattern is transferred to the intermediate germanium layer by reactive ion etching in carbon tetrafluoride $\mr{CF_4}$.
Next, inside the same chamber, the pattern is transferred to the bottom P(MMA-MAA) layer by anisotropic oxygen plasma etching.
Immediately after this, an undercut profile is created by etching this layer with isotropic oxygen plasma.
After this, we metallize the structure, first by depositing $20\,\mr{nm}$ of aluminium by e-beam evaporation at an angle of $15.2^\circ$ to form the leads.
Inside the evaporation chamber, this layer is oxidized with $2.2\,\mr{mbar}$ of oxygen for $2$ minutes.
To create the island and finally form the SINIS device, $30\,\mr{nm}$ of copper are deposited at an angle of $-14.8^\circ$.
The chip is then bathed in acetone for removing excess metal and remaining P(MMA-MAA).

\subsection{Measurements}
After fabrication, the chip is cleaved to fit in a custom made sample holder to which one device is bonded.
This sample holder has been modified so that surface mounting inductors and capacitors form bias tees between DC and radio-frequency (RF) inputs, which are then connected to the sample gate and source electrodes by aluminum wires.
The remaining drain electrode is connected to a DC line only.
DC levels of the applied signals are injected through the bias tee inductors while RF signals are injected through the bias tee capacitors.
The sample holder is then attached to the mixing chamber of a custom made dilution refrigerator with a base temperature of $\sim 100\,\mr{mK}$.
We apply DC signals through cryogenic lines composed of resistive twisted pairs running from room temperature down to the fridge $1\,\mr{K}$ flange and nearly $1\,\mr m$ of Thermocoax cable down to the mixing chamber.
RF signals were applied through the lines consisting of stainless steel coaxial cable installed between the room temperature (top) and $4.2\,\mr{K}$ flanges, a $20\,\mr{dB}$ attenuator at this temperature, followed by a feedthrough into the inner vacuum can inside which a NbTi coaxial cable follows from the $1\,\mr{K}$ flange down to the RF input of the holder.
Additionally, at room temperature, $40\,\mr{dB}$ attenuation is applied to the line carrying the source-drain RF bias signal and a further $20\,\mr{dB}$ attenuator is connected to the gate RF line.
We generate DC and AC signals by programmable voltage sources and waveform generators, respectively.
To measure current, we use a digital multimeter for reading out the voltage in the output of the transimpedance current amplifier (FEMTO Messtechnik, model LCA-2-10T) connected to the device drain electrode through a DC line.
To ensure proper synchronization and phase shift between the bias and gate RF signals, both were generated by of a 2-channel arbitrary waveform generator (Keysight, model 33522B).
Each measurement of the pumped current was iterated typically 15 times and later averaged subtracting those repetitions during which a charge offset jump had occurred. 
Then, the offset of the current amplifier was subtracted by comparing the curves measured with biases of equal magnitude and opposite polarity.

\section{Device characterization and new proposal}\label{Sec:Char}
Figure \ref{f1}(b) shows the device DC current-voltage characteristics, which we measure by sweeping the DC gate voltage through about two gate periods for each DC bias voltage (green vertical lines).
We calculate the maximum and minimum current using a Markovian equation (magenta lines, see Appendix for details on this model) with the following parameters: $\Delta =210\,\mr{\mu eV}$, $\ec=2.48\Delta$, total normal-state tunnel resistance $\rt=4.53\,\mr{M\Omega}$, ratio between left and right junction tunnel resistances $r=0.15$ and the Dynes parameter~\cite{Dynes1978} $\eta =3.5\times 10^{-4}$.
The IV curves simulated with these parameters agree well with the measured data.
The superconducting gap of the leads creates a bias voltage zone inside of which, ideally irrespective of the gate voltage, no current flows, see Fig.~\ref{f1}(b).
For biases $2\Delta<\vob<2\Delta+2\ec$ current is suppressed only for certain gate voltages.
These features create a diamond-like structure in the bias voltage-gate voltage parameter space inside of which the island charge remains stable.
Figure~\ref{f1}(c) depicts two of these (only the upper half is shown) for states with zero and one extra electron in the island for a device with $r=1$, notice how the diamonds overlap.
These zones are bounded by the tunnelling thresholds, which, if crossed, trigger a single-electron tunnelling event between the island and one of the leads, either left (L) or right (R).
These are defined by
\begin{equation}
\Delta=\pm 2\ec\left(n-\nng\pm 0.5\right)\pm eV_\mr{b,L/R}\equiv \delta\epsilon^\pm_\mr{L/R}.
\label{Eq:threshold}
\end{equation}
We adopt the convention that positive biasing is from left (L) to right (R), hence $V_\mr{b,L}=\kappa_\mr L\vb$ and $V_\mr{b,R}=-\kappa_\mr R\vb$, where $\kappa_i$ is the ratio between the junction $i$ capacitance and the total capacitance.
Additionally, $n$ is the initial island charge state.
In Eq.~\eqref{Eq:threshold} the plus signs indicate a tunnelling event into the island and minus signs indicate events out of the island.
The thresholds for $n=0,1$ are depicted in Fig.~\ref{f1}(c), see its caption for further explanation.
The key to the turnstile operation is to follow a path that crosses a threshold for electron tunnelling into the island through one junction and then crosses another for a tunnelling out of the island through the opposite junction without leaving the stability region.
The most basic trajectory is depicted in Fig.~\ref{f1}(c) as path (1), henceforth called flat driving.
Following this path back and forth with frequency $f$ generates current $I=ef$~\cite{Pekola2007}.
However, in Fig.~\ref{f1}(c) the curve designated as (2) is also suitable for generating DC currents.
Such a path can be realized by adding an extra modulation to the bias such that
\begin{equation}
\begin{split}
\nng&=\nog+\dfrac{\ag}{2}\sin{\left(\omega t\right)},\\
\vb&=\vob+\dfrac{\ab}{2}\cos{\left(2\omega t\right)},
\end{split}
\label{eq:protocol}
\end{equation}
with $\omega =2\pi f$. Considering time $t$ as a parameter, Eq.~\eqref{eq:protocol} verifies $\vb=\vob+\ab/2-4\left(\nng-\nog\right)^2\ab/\ag^2$, that is, a parabola with negative concavity.
Here, $\ab$ is the peak-to-peak (pp) amplitude of the bias signal, $\ag$ the pp amplitude of gate signal, $\nng=\cg\vg/e$ with $\cg$ the capacitance between the gate and the island and $\vg$ the voltage applied to the gate, in our device $\cg=7.28\,\mr{aF}$.
In practice, $\nog$ is set to the gate open position at the bottom of the parabola given by $\nog^\mr{open}=\left(\vob-\ab/2\right)\left(r-1\right)/\left[4\ec\left(r+1\right)\right]+0.5$, assuming $r=\kappa_\mr{R}/\kappa_\mr{L}$, which is a good approximation in this case.
However, the choice of $\nog$ only affects the width of the current plateaus against $\ag$ and whether or not certain ones appear.
Notice that this path can be reduced to the flat one by setting $\ab=0$.
\begin{center}
\begin{figure}
\includegraphics[scale=0.95]{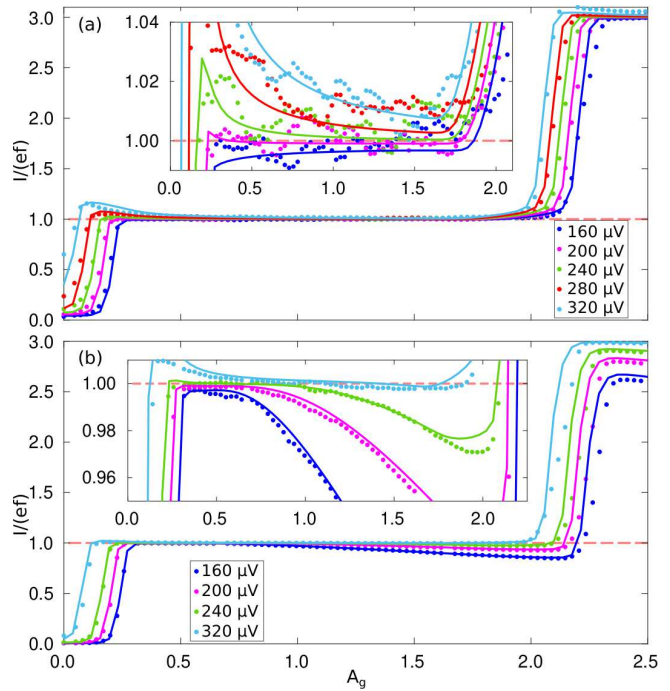}
\caption{(a) Typical single-electron current plateaus at $f=1\MHz$ with only DC bias.
The legend shows the corresponding DC bias voltage $\vob$.
Dots are measured data, solid lines are simulations.
Inset: zoom-in around $I/(ef)=1$ showing the accuracy of the measured single-electron current and of our modelling.
(b) As in panel (a) for $f=5\MHz$.
Note how the plateaus bend down which is a clear signature of back-tunnelling.}
\label{f2}
\end{figure}
\end{center}

The results of applying the flat driving are depicted in Fig.~\ref{f2} as dots along with calculations based on the same Markovian model used in the DC characterization and the same parameters, as solid lines.
Our calculations carefully follow the dependence of the current against the driving amplitude for several biases, capturing deviations from the quantized current.
Notice that for $f=1\MHz$ (Fig.~\ref{f2}(a)) $I\sim ef$ in a broad $\ag$ interval (see inset) hence forming a clear plateau for most of the applied bias voltages except when excess events start to appear (for which $I>ef$).
In contrast, for $f=5\MHz$ (Fig.~\ref{f2}(b)) these current plateaus bend down in a wide range of gate amplitudes such that $I<ef$ (see inset) and the accuracy decreases with increasing $\ag$ until the next current plateau is reached.
This occurs because when the driving amplitude increases, the parametric curve it traces crosses additional tunnelling thresholds before reaching the next charge state.
These thresholds allow tunnelling events against the bias direction, referred to as back-tunnelling events, which limit accuracy in Fig.~\ref{f2}(b).
Such processes will only take place when the island has not been filled or emptied by the desired (forward-)tunnelling, which happens when the corresponding forward-tunnelling rates are comparable to the inverse of the time elapsed between the crossings of the favorable and adjacent unfavorable thresholds, we designate this interval duration by $\delta t$.
For continuous drivings the extent of this time interval decreases with increasing frequency.
There is an additional proportionality on $\vob$ whose form depends on the specific path and waveform, this is clearly shown in Fig.~\ref{f2}.
Thus, in order to generate accurate single-electron current, one should increase $\delta t$.
We propose that applying the parabolic driving helps in this regard since it reshapes the quantities $\delta\epsilon^\pm_ \mr{L/R}$ so that the desired events cross the threshold with a lower rate and the undesired ones do this at a higher rate, thus increasing $\delta t$.

\section{Effect of the parabolic driving on the system dynamics}\label{Sec:Effect}
\begin{center}
\begin{figure}
\includegraphics[scale=0.61]{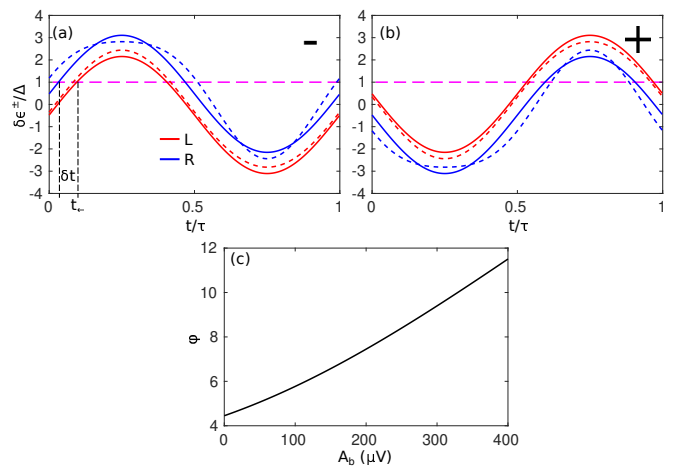}
\caption{(a) Time evolution of the system energy difference for single-electron tunnelling out of the island in the flat protocol (solid lines) and in the parabolic protocol (dashed lines).
Red indicates processes through the left junction, blue through the right, the magenta dashed line designates the tunnelling thresholds, $\tau$ is the period of the driving signal.
(b) As in (a) for single-electron tunnelling into the island.
(c) Integral of the forward tunnelling rate against bias amplitude in the cases of panels (a) and (b).
}
\label{f5}
\end{figure}
\end{center}

We clarify such proposition in this Section by explaining how the parabolic protocol affects the island chemical potential evolution and what is its impact on back-tunnelling events based on the Markovian model described in the Apendix.
As shown there, the time evolution of the system is contained in $\delta\epsilon_\LR^\pm$ (Eq.~\eqref{Eq:threshold}) through $\nng$ and $\vb$.
These quantities control the tunnelling rates (Eq.~\eqref{eq:rate}), which, ideally, become non-zero only when $\delta\epsilon_\LR^\pm \geq\Delta$, that is, when the energy thresholds defining the stability diagram of an SET are crossed.
We plot in Figs.~\ref{f5}(a)-(b) the evolution of $\delta\epsilon_\LR^\pm$ within one driving period $\tau$, where each panel corresponds to one kind of the tunnelling process (either out of, Fig.~\ref{f5}(a), or into the island, Fig.~\ref{f5}(b)) for both junctions.
In the case of the flat driving (solid lines), the evolution is entirely controlled by the gate signal $\nng$ and accordingly adopts its shape, we present the curves for $\ag=1.06$ and $V_\mr{0b}=200\,\mr{\mu V}$.
For the parabolic driving (dashed lines), this evolution is also controlled by $\vb$, hence $\delta\epsilon_\LR^\pm$ adopts a different time evolution as compared to the flat driving; in Fig.~\ref{f5} we used $A_\mr{b}=240\,\mr{\mu V}$.
The tunnelling thresholds are depicted as magenta dashed lines in Fig.~\ref{f5}.
Notice that the curve for the desired tunnelling events ($\delta\epsilon_\mr{R}^-$ and $\delta\epsilon_\mr{L}^+$, in our particular case) crosses the threshold before the undesired ones ($\delta\epsilon_\mr{L}^-$ and $\delta\epsilon_\mr{R}^+$, in this case), whose crossings are only possible if the modulation amplitude is large enough.
Hence, ideally, the former events will happen first making the latter ones impossible.
However, if the time elapsed between the crossings of the two thresholds ($\delta t$ in Fig.~\ref{f5}(a)) is too short, then the favourable event may not take place before the threshold crossing of the undesired event, therefore allowing the latter to occur.
In the case of Fig.~\ref{f5}, there is clear back-tunnelling for the flat driving when $f=1/\tau =5\,\mr{MHz}$, see Fig.~\ref{f2}(b).
Notice how the use of the parabolic driving reshapes the curves and extends $\delta t$.

\begin{figure*}
\includegraphics[scale=0.96]{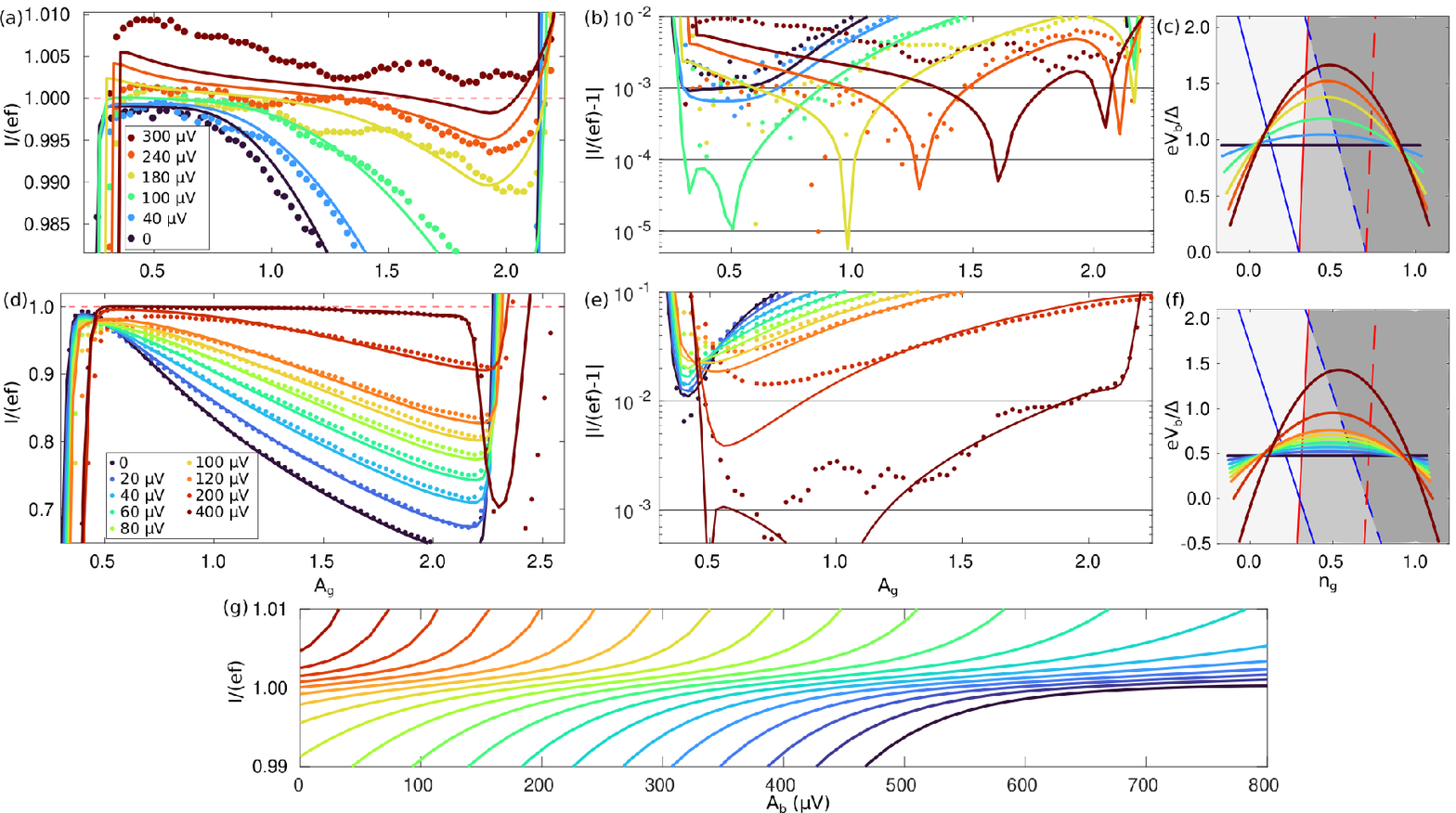}
\caption{(a) Current plateaus using the parabolic driving at $f=5\MHz$ and $\vob=200\uv$, the legend designates the used pp bias amplitude.
Dots represent measurements, solid lines simulations.
(b) Deviation from ideal $\left|I/\left(ef\right)-1\right|$ of data and calculations of panel (a) with the same legend.
(c) Depiction of the used protocols in the $\nng-\vb$ space.
Colors correspond to the curves of panel (a) and (b).
The stability diagram corresponds to the estimated parameters of the measured device.
It has been zoomed-in to the intersection between charge states for clarity.
Tunnelling thresholds correspond to the convention of Fig.~\ref{f1}(c).
(d) Measured current plateaus using the proposed protocol at $f=5\MHz$ and $\vob=100\uv$, the legend designates the used pp bias amplitude.
(e) As in (b) corresponding to panel (d).
(f) As in (c) corresponding to panel (d) and (e).
(g) Current calculated at the plateau ($\ag=1.2$) as a function of the pp bias amplitude of the parabolic protocol, curves from right to left go from $\vob=0$ to $360\uv$ in steps of $20\uv$.
}
\label{f3}
\end{figure*}

We show, based on a simplification of the model described in the Appendix, that by extending this time interval back-tunnelling is suppressed.
First, we simplify the model by assuming zero temperature, $T_\mr{N}=T_\mr{S}=0$, and perfect junctions, $\eta =0$.
Under these conditions, the tunnelling rates Eq.~\eqref{eq:rate} become
\begin{equation}
\Gamma\left(\delta\epsilon_\LR^\pm\right)=\dfrac{1}{e^2R_\LR}\sqrt{\left(\delta\epsilon_\LR^\pm\right)^2-\Delta^2},
\end{equation}
which is valid for $\delta\epsilon_\LR^\pm\geq \Delta$.
In our particular case, the right junction, in so far as the less transparent one, sets the lower bound for back-tunnelling, because tunnelling rates through this junction are lower.
Therefore we focus on the events occurring through the right junction.
We make a further approximation by stating that the probability of having one electron in the island at the time when the back-tunnelling process energy crosses the threshold $t_\leftarrow$ (that is, when $\delta\epsilon_\mr{L}^-=\Delta$, see Fig.~\ref{f5}(a)) is~\cite{Pekola2022}
\begin{equation}
p_1(t_\leftarrow)\approx \exp{\left(-\int_{\delta t}\Gamma\left(\delta\epsilon_\mr{R}^-\right)\, dt\right)}.
\end{equation}

Having $p_1(t_\leftarrow)\ll 1$ means that the island has already been emptied before back-tunnelling becomes energetically favorable, while $p_1(t_\leftarrow)\lesssim 1$ means that the island is not necessarily empty at $t_\leftarrow$ and hence it will likely be emptied through back-tunnelling.
Therefore, the larger the quantity $\varphi=\int_{\delta t}\Gamma\left(\delta\epsilon_\mr{R}^-\right)dt$, the less likely back-tunnelling.
In the case of Fig.~\ref{f5}(a), $\varphi_\mr{flat}\approx 4.45$ for the flat driving while $\varphi_\mr{par}\approx 8.20$ for the parabolic one.
We have then, $\varphi_\mr{par}>\varphi_\mr{flat}$.
Hence, applying the parabolic driving reduces the likelihood of back-tunnelling events.
Fig.~\ref{f5}(c) shows that $\varphi$ grows with $A_\mr{b}$, therefore a larger bias amplitude suppresses back-tunnelling more efficiently.

\section{Results and discussion}\label{Sec:Results}
\subsection{Experimental results}
We prove this experimentally by presenting the measured results for $f=5\MHz$ in Fig.~\ref{f3} where back-tunnelling has a strong influence on the current plateau.
Fig.~\ref{f3}(a) shows the results of applying the parabolic driving with $\vob=200\uv$ for several bias amplitudes $\ab$ as dots.
Note that the Markovian model agrees well with the measured current after applying a correcting factor of two to the amplitude of the AC signal delivered to the source electrode, which we deem as a result of the RF line transmittance, see the solid lines in Fig.~\ref{f3}(a).
From these curves it is evident that the overall effect of the driving is to increase the current in the back-tunnelling affected regime, a clear signature of back-tunnelling suppression.
These detrimental processes are replaced by forward-tunnelling events which keep the current closer to the desired value $I=ef$.
Fig.~\ref{f3}(b) shows the deviation from ideal $\left|I/\left(ef\right)-1\right|$ of data in Fig.~\ref{f3}(a), notice that the flat driving at $\vob=200\uv$ gives a current accuracy of $\sim 10^{-3}$ at best.

Instead, using a pp bias amplitude of $100\uv,\,180\uv$ and $240\uv$ lowers the deviation to the $10^{-4}$ level.
A depiction of the paths followed in the stability diagram of the measured device can be seen in Fig.~\ref{f3}(c).
From there it is clear that for larger $\ab$ the path length between the threshold crossings for desired events and for undesired ones increases.
For $\vob=100\uv$ the improvement is more drastic, see Figs.~\ref{f3}(d) and (e).
The minimum deviation achieved with the flat driving is $>10^{-2}$ while this is on the level of $10^{-3}$ and even below for a narrow interval of gate amplitudes for $\ab=400\uv$, Fig~\ref{f3}(f) shows the followed paths.
Furthermore, for most of the plateau the error even exceeds $10^{-1}$ in the first case while in the second one the error remains below $10^{-2}$ for a wide $\ag$ interval.
This demonstrates that even for the conditions in which accuracy is otherwise poor we can increase it by one order of magnitude or more using the extra bias modulation.
In Fig.~\ref{f3}(d) we can see that the maximum current (against $\ag$) decreases with increasing $\ab$ before starting to increase.
Interestingly, this happens in the gate amplitude interval where back-tunnelling is not energetically favorable.
According to our model, this effect is due to an interplay between the reshaping that the protocol causes on $\delta\epsilon^\pm_\mr{L/R}$ and the sub-gap states in the superconducting leads (leakage).
Because of these, tunnelling is possible even when $\delta\epsilon^\pm_\mr{L/R}<\Delta$, though with lower rates.
In general, the parabolic driving increases the maxima of these quantities for undesired tunnelling while decreasing them for desired events. 
Therefore, back-tunnelling sub-gap rates increase while lowering the forward-tunnelling ones, and current decreases.
This imbalance is accentuated with increasing $\ab$ and is only important as long as the forward-tunnelling thresholds are not crossed.
Such an effect is minimized with better junctions having a lower Dynes parameter.
To understand how the current at the plateaus varies with $\ab$, we plot in Fig.~\ref{f3}(g) the former against the latter.
These curves have been calculated with a constant $\ag=1.2$.
We see that even for $\vob=0$ our measured protocol can recover the expected current for large enough $\ab$.
However, the higher the DC bias, the more sensitive the current becomes to the bias amplitude.
In any case the current tends to converge around the ideal value, although not with the same robustness as against $\ag$ or $\vob$.
Ideally, one expects the current not to depend very strongly on $\ab$ around $ef$ in order to have a good standard, therefore one would expect to see plateaus against this parameter.

\subsection{Future prospects}
\begin{figure*}
\includegraphics[scale=1.08]{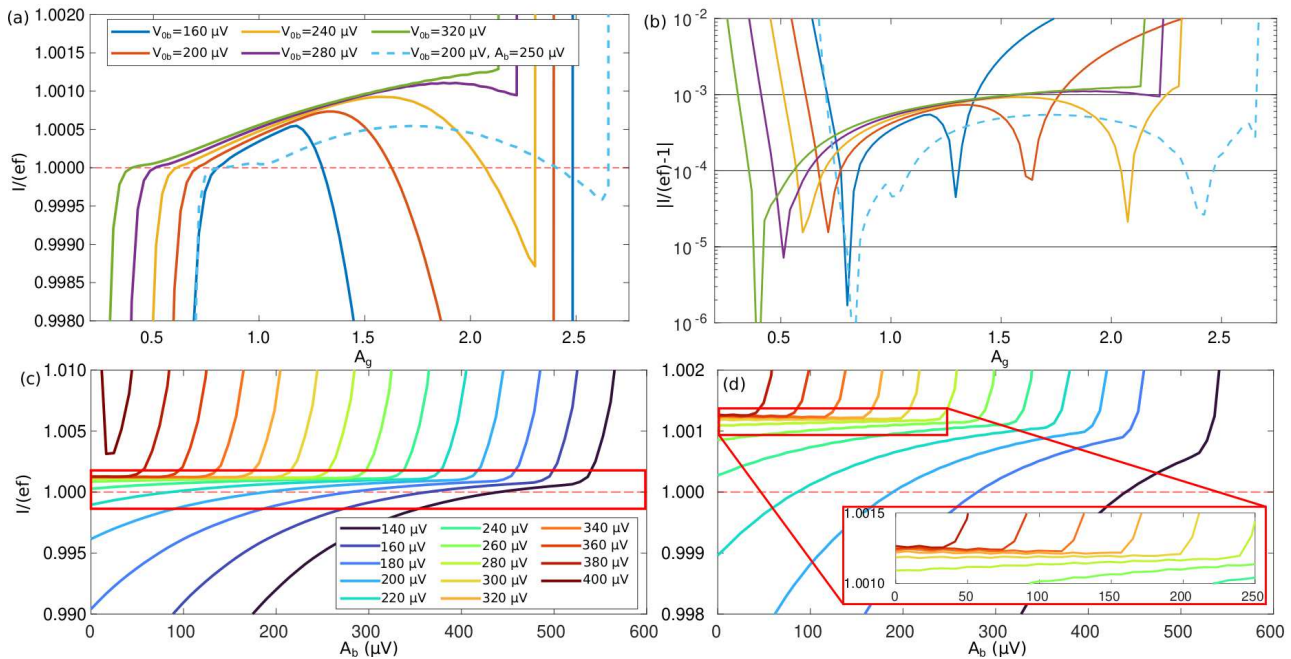}
\caption{(a) Simulated current plateaus.
The legend shows the biasing conditions for the flat protocol.
The dashed line was calculated using the parabolic protocol with the amplitude given in the legend.
(b) Deviation of current of panel (a) with the same legend.
(c) Calculated current at the plateau $(\ag=2.0)$ against the protocol amplitude $\ab$. Legend indicates the used $\vob$
(d) Zoom-in of the enclosed area of panel (c).
Inset: zoom-in of the enclosed area.
}
\label{f6}
\end{figure*}
We use our theoretical model to evaluate how promising the parabolic driving is.
For this, we simulate the turnstile operation for a device with $\Delta =200\,\mr{\mu eV}$, $\ec=1.2\Delta$, $\rt=200\,\mr{k\Omega}$, $\eta =10^{-6}$ and $r=1$.
These parameters ensure that the turnstile produces accurate single-electron currents for driving at high frequency.
Furthermore, we set the island temperature to $T_\mr{N}=10\,\mr{mK}$ and the leads temperature to $T_\mr{L/R}=100\,\mr{mK}$, which corresponds to a good quasiparticle relaxation.
These are realistic device characteristics that can be routinely obtained with dedicated fabrication and setup conditioning \cite{Pekola2010,Knowles2012}.
We show in Fig.~\ref{f6} the calculated current at operation frequency $f=240\MHz$ for which, ideally, $I\approx 38.45\,\mr{pA}$.
It is clear that with the flat driving, even though certain biases are subject to back-tunnelling except for $\vob=280\uv$ and $\vob=320\uv$, accurate single-electron current emission is achievable with plateaus ranging between $10^{-4}$ and $10^{-3}$ deviation, see Figs.~\ref{f6}(a) and (b).
It is also evident that the application of the parabolic driving with $\vob=200\uv$ and $\ab=250\uv$ increases the accuracy of the current compared to the flat driving when the same DC bias level is applied since it dramatically reduces back-tunnelling.
Moreover, the current produced with the new protocol is clearly more accurate than any other presented in Fig.~\ref{f6}(a).
In fact, although deviations go even to $10^{-5}$ for the flat driving, we get a broader region for which the error falls below $10^{-3}$ and $10^{-4}$ when using our new approach.
Additionally, the implementation of our method improves the flatness of the plateau.
Fig.~\ref{f6}(c) shows the behavior of the current against $\ab$ for several values of $\vob$ at $\ag=2$.
We can see that the robustness of the plateaus improves compared to Fig.~\ref{f3}(g).
Furthermore, for certain DC bias values the current remains very much constant (see Fig.~\ref{f6}(d) and inset) although for these there is generally no discernible improvement in current accuracy when employing the new method ($\ab\neq 0$).
Here, as well as in Fig.~\ref{f6}(a) we ascribe the reduction of excess current to an enhancement of back-tunnelling-induced leakage by the parabolic protocol as previously explained.

Further implementations of SET driving with bias modulation could include a phase shift of this signal by $\pi$, inverting the parabola.
This would decrease the time-window for forward-tunnelling-only, however, it could also reduce the current leakage.
We would like to stress that all in all the use of any kind of modified driving protocol involving bias modulation alters the rate at which $\delta\epsilon_\mr{L/R}^\pm$ passes the tunnelling thresholds while keeping the operation frequency.
Thus, these strategies can be used to improve the accuracy of frequency-to-power conversion of Ref. \citenum{Marin-Suarez2022} by reducing the threshold passing rate of the energies~\cite{Pekola2022} while keeping the magnitude of generated power constant.
Apparently, additional improvements are needed in order to attain the accuracy of $10^{-7}$ at the output currents of $\sim 100\,\mr{pA}$ demonstrated in the semiconductor electron pumps with a tunable barrier~\cite{Stein2015,Stein2016}.
These may include a better protection from the environmental photons that promote leakage while allowing proper bias driving.
Lower tunnel resistances are also necessary while $\ec>\Delta$ is kept so that the unwanted tunnelling events due to Andreev reflection are suppressed.

\section{Conclusions}\label{Sec:Conclusions}
In summary, we have, for the first time, tested a SET driving involving bias modulation in a SINIS device added to the commonly used periodic gate modulation.
This second signal applied to the source electrode of the transistor has a frequency twice that of the gate driving, and
an appropriate phase-locking with respect to it. Furthermore, we provided evidence of the advantage of using this scheme against the usual one under diverse conditions.
We conclusively showed that the main effect of this new approach is the suppression of the tunnelling events against the bias.
These events compromise the accuracy of single-electron transport at high gate driving frequencies.
A Markovian description of the system, which accurately models the measured currents, supports our arguments on the superiority of the method employed here.
We also showed that this driving can be employed with noticeable improvement for a broad interval of source modulation amplitudes.
Furthermore, based on the theoretical model, we made clear that our approach holds the promise of improving the accuracy in devices with optimized parameters at larger single-electron currents with respect to the usual driving.
This work opens the path for testing new SET protocols other than gate driving alone.
Many more driving methods can be implemented by modifying the used waveforms as well as the frequency and phase shift of the bias modulation.

\section*{Acknowledgments}
We acknowledge fruitful discussions with S. Kafanov.
M.M.-S., J.T.P. and J.P.P. acknowledge support from Academy of Finland under grant number 312057.
Yu.A.P. acknowledges support from the QSHS project ST/T006102/1 funded by STFC.

\section*{Appendix: Model description and numerical simulations}
\setcounter{equation}{0}
\renewcommand\theequation{A\arabic{equation}}
\subsection{General model}
We described the SINIS transistor dynamics using the Markovian model described in the Supporting Information of Ref.~\onlinecite{Marin-Suarez2020}.
Here we reproduce this explanation with modifications pertinent for the present study.

We model the dynamics of a single-electron transistor from a stochastic master equation for the number of excess charges in the island $n$.
The probability of having $n$ excess electrons (also referred to as the state $n$), $p\left(n\right)$, evolves as \cite{Kulik1975,Likharev1985,Averin1986,Saira2013}
\begin{equation}
\dfrac{d}{dt}p\left(n,t\right)=\sum_{n'\neq n}{\gamma_{n'n}p\left(n',t\right)-\gamma_ {nn'}p\left(n,t\right)},
\label{e1}
\end{equation}
where $\gamma_{nn'}$ is the total transition rate of the system for going from state $n$ to $n'$.
In the specific case of an SET, the transition rates are given by $\gamma_{nn'}=\Gamma_{n\rightarrow n'}^\mr{L}+\Gamma_{n\rightarrow n'}^\mr{R}$.
Here, $\Gamma_{n\rightarrow n'}^\LR$ is the transition rate between individual charge states $n$ and $n'$ due to tunnelling through either left (L) or right (R) junction.

We restrict the model to single-electron tunnelling and two-electron Andreev reflection.
In a NIS junction the rates for these processes are given by
\begin{widetext}
\begin{equation}
\Gamma^\LR_{n\rightarrow n\pm 1}\left(\delta\epsilon_\mr{L/R}^\pm\right)=\dfrac{1}{e^2\rt}\int dE\,{n_\mr{S}\left(E-\delta\epsilon_\mr{L/R}^\pm\right)f_\mr{N}\left(E\right)\left[1-f_\mr{S}\left(E-\delta\epsilon_\mr{L/R}^\pm\right)\right]},
\label{eq:rate}
\end{equation}
\end{widetext}
for single-electron tunnelling, and
\begin{widetext}
\begin{equation}
\begin{split}
&\Gamma^\LR_{n\rightarrow n\pm 2}\left(\delta\epsilon_\mr{L/R}^{\pm 2}\right)=\\
&\dfrac{\hbar\Delta^2}{16\pi e^4\rt^2\mathcal{N}}\int dE\, f_\mr{N}\left(E+\delta\epsilon_\mr{L/R}^{\pm 2}/2\right)f_\mr{N}\left(-E+\delta\epsilon_\mr{L/R}^{\pm 2}/2\right)\left|a\left(E+\ec-i\xi/2\right)+a\left(-E+\ec-i\xi/2\right)\right|^2, \label{e9}
\end{split}
\end{equation}
\end{widetext}
for Andreev tunnelling.
For the case of Eq.~\eqref{eq:rate} $\delta\epsilon_\mr{L/R}^\pm$ is given in Eq.~\eqref{Eq:threshold}, and for Eq.~\eqref{e9}
\begin{equation}
\delta\epsilon_\mr{L/R}^{\pm2} =\pm 4\ec\left(n-\nng\pm 1\right)\pm 2eV_\mr{b,L/R}.
\end{equation}
Here $V_\mr{b,L}=\kappa_\mr{L}\vb$ and $V_\mr{b,R}=-\kappa_\mr{R}\vb$, $\kappa_\LR$ is the ratio between the junction capacitance and the total capacitance, $\vb$ is the bias voltage applied between the source-drain leads of the transistor, $n$ is the initial island excess charge, $n_\mr{g}$ is the charge number induced by the gate voltage, $E_\mr{c}$ is the charging energy and $+\,(-)$ designates tunnelling to (from) the island.

In Eqs.~\eqref{eq:rate} and \eqref{e9}, $\Delta$ is the superconducting gap of the leads, $\rt$ is the tunnel resistance of the junction involved in the event and $\mathcal{N}$ is the number of conduction channels which can be written as $A/A_{\mathrm{ch}}$, with $A$ being the junction area ($\sim 50\,\mr{nm}\times 60\,\mr{nm}$ for the present device) and $A_{\mr{ch}}$ is the area of an individual channel, estimated to be $30\,\mr{nm^2}$, although this precise value does not affect the results of the model in the present case since $\ec>\Delta$~\cite{Aref2011}.
The term $\xi$ takes into account the energy of the intermediate (single-electron tunnelling) state which has a finite lifetime and correspondingly can be calculated as $\hbar\sum_\pm\Gamma_{n\rightarrow n\pm 1}$ ~\cite{Averin2008}, for the present case we use $\xi/\Delta =10^{-5}$.
For our purposes, the exact value of this quantity does not have an impact on the value of the rates~\cite{Maisi2014a}.
Additionally, $f_\mr{N}$ is the Fermi-Dirac distribution function for electrons in the normal-metal island, $f_\mr{S}$ is that for the superconducting lead involved in the tunnelling event and $n_\mr{S}$ is the superconducting density of states given by~\cite{Bardeen1957}.
\begin{equation}
n_\mr{S}\left(E\right)=\left|\mathfrak{Re}\left(\dfrac{E/\Delta+i\eta}{\sqrt{\left(E/\Delta+i\eta\right)^2-1}}\right)\right|.
\label{e11}
\end{equation}
Here $\eta$ is the Dynes parameter that helps model sub-gap leakage~\cite{Dynes1978}. Finally,
\begin{equation}
a\left(x\right)=\dfrac{1}{\sqrt{x^2-\Delta^2}}\ln\left(\dfrac{\Delta-x+\sqrt{x^2-\Delta^2}}{\Delta-x-\sqrt{x^2-\Delta^2}}\right).
\end{equation}

In our work we have measured the SET in two different regimes, namely the DC and turnstile operations.

\subsection{DC operation}
For calculating the current through the SET as response to applied DC bias ($\vb$) and gate voltages ($\vg$) through $\nng =C_\mr{g}\vg/e$, we solve Eq.~\eqref{e1} in steady state, that is, $dp\left(n,t\right)/dt=0$.
Since the charge states $n$ are discrete, one can express Eq. \eqref{e1} as a matrix equation
\begin{equation}
A\mathbf{p}=\mathbf{0},
\label{eq:dc_eq}
\end{equation}
where $p_n=p\left(n\right)$, $A_{nn}=-\sum_{n'\neq n}\gamma_{nn'}$ and $A_{nn'}=\gamma_{nn'}$ for $n\neq n'$.
Hence, the steady-state probability of having $n$ electrons in the island is found through calculating the null space of $A$.

In order to get accurate results in the DC regime, heating of the island due to non-zero power dissipation has to be taken into account.
To do this, we need to calculate the total power transferred to the normal-metal island in the steady-state regime by proposing a vector $\mathbf{q}$ such that $q_n=\dot{Q}^\mathrm{N}_{n\rightarrow n+1}+\dot{Q}^\mathrm{N}_{n\rightarrow n-1}$ with $\dot{Q}^\mathrm{N}_{n\rightarrow n\pm 1}=\dot{Q}^\mathrm{N,R}_{n\rightarrow n\pm 1}+\dot{Q}^\mathrm{N,L}_{n\rightarrow n\pm 1}$, where
\begin{widetext}
\begin{equation}
\dot{Q}^\mr{N,\LR}_{n\rightarrow n\pm 1}\left(\delta\epsilon_\mr{L/R}^\pm\right)=\dfrac{1}{e^2\rt}\int dE\,{En_\mr{S}\left(E-\delta\epsilon_\mr{L/R}^\pm\right)f_\mr{N}\left(E\right)\left[1-f_\mr{S}\left(E-\delta\epsilon_\mr{L/R}^\pm\right)\right]},
\end{equation}
\end{widetext}
that is, the power transferred by the electron tunnelled in a single event.
Also, $\delta\epsilon_\mr{L/R}^\pm$ is the related energy cost of single-electron tunnelling defined by Eq. \eqref{Eq:threshold}.
Then the total power transferred to the island is $\dot{Q}=\mathbf{q}\cdot\mathbf{p}$.
This power is dissipated by electron-phonon interaction~\cite{Wellstood1994} creating an equilibrium state in which the phonon temperature $T_0$ (which is taken as the temperature of the bath) differs from the temperature of the island electron system $T_\mr{N}$.
These temperatures depend on the power dissipated as $\dot{Q}_\mathrm{e-ph}=\mathcal{V}\Sigma\left(T_\mathrm{N}^5-T_\mr{0}^5\right)$, where $\mathcal{V}$ is the volume of the island and $\Sigma$ is the electron-phonon coupling constant ($\approx 8.4\times 10^9\,\mr{WK^{-5}m^{-3}}$ for copper, which was obtained in earlier experiments~\cite{Marin-Suarez2022}).
An additional power transfer due to Andreev reflection is considered in the form of Joule heat, that is, $\dot{Q}_\mr{A}=\left\langle I_\mathrm{A}\right\rangle V_\mathrm{b}$~\cite{Rajauria2008}, where $I_\mathrm{A}$ is the current due only to Andreev events and $V_\mathrm{b}$ is the applied bias voltage.
Finally, the heat balance is $\dot{Q}_\mathrm{e-ph}=\dot{Q}+\dot{Q}_\mathrm{A}$, in the present case $\dot{Q}_\mr{A}\sim 0$.
Notice that the transition rates depend also on $T_\mr{N}$.
Hence, this quantity has to be solved for self-consistently to satisfy Eq.~\eqref{eq:dc_eq} and the heat balance.

Once $T_\mr{N}$ is determined with sufficient accuracy (for this we set a tolerance of $0.01\,\mr{mK}$), Eq.~\eqref{eq:dc_eq} can be solved for $\mathbf{p}$.
Then, the current through the SET $I$ can be calculated as 
\begin{equation}
\begin{split}
I&=\mathbf{b}\cdot\mathbf{p},\,\mr{where}\\
b_n=e&\left(\Gamma_ {n\rightarrow n+1}^\mr{L}-\Gamma_{n\rightarrow n-1}^\mr{L}\right)+\\
2e&\left(\Gamma_ {n\rightarrow n+2}^\mr{L}-\Gamma_{n\rightarrow n-2}^\mr{L}\right).
\end{split}
\label{eq:current}
\end{equation}

\subsection{Turnstile operation}
For modelling the turnstile operation we cannot use Eq.~\eqref{eq:dc_eq}.
Instead we have
\begin{equation}
\dfrac{d}{dt}\mb{p}\left(t\right)=A\left(t\right)\mathbf{p}\left(t\right).
\label{eq:turn_eq}
\end{equation}
The time dependence of matrix $A$ stems from the time dependence of $\delta\epsilon_\mr{L/R}^\pm$, which, in turn, is contained in $\nng$ for the flat driving and additionally in $\vb$ for the parabolic driving, see Eqs.~\eqref{eq:protocol} in the main text.

To calculate the current, we propose an evolution equation for the state averaged charge in the island $\left\langle q\right\rangle_\mr{s}$~\cite{Saira2013}
\begin{equation}
\dfrac{d}{dt}\begin{bmatrix}
\mb{p}\left(t\right) \\
\left\langle q\right\rangle_\mr{s}\left(t\right)
\end{bmatrix}
= \begin{bmatrix}
A\left(t\right) & \mb{0} \\
\mb{b}\left(t\right)^\mr{T} & 0
\end{bmatrix}
\begin{bmatrix}
\mb{p}\left(t\right) \\
\left\langle q\right\rangle_\mr{s}\left(t\right)
\end{bmatrix}.
\label{eq:turn_eq_extended}
\end{equation}
Notice that Eq.~\eqref{eq:turn_eq_extended} includes Eq.~\eqref{eq:current}.

The solutions of Eq.~\eqref{eq:turn_eq_extended} are given by
\begin{equation}
\begin{bmatrix}
\mb{p}\left(t\right) \\
\left\langle q\right\rangle\left(t\right)
\end{bmatrix}
=\exp{\left(\int_0^t{dt'\begin{bmatrix}
A\left(t'\right) & \mb{0} \\
\mb{b}\left(t'\right)^\mr{T} & 0
\end{bmatrix}}\right)}
\begin{bmatrix}
\mb{p}\left(0\right) \\
\left\langle q\right\rangle\left(0\right)
\end{bmatrix}.
\label{eq:solution}
\end{equation}

In the specific operation of the single-electron turnstile, one assumes periodic boundary conditions for $\mb{p}$ with the same period as $\nng$, the driving signal.
This is reasonable since the tunnelling rates have the same periodicity as this signal.
Therefore, one can approximate the integral by discretizing the driving cycle of period $\tau$ in $m$ intervals of size $\delta\tau=\tau/m$.
At the end of the period the exponential is then
\begin{equation}
\tilde{U}(\tau)=\prod_{k=1}^m{\exp\left({\delta\tau\begin{bmatrix}
A\left(t_k\right) & \mb{0} \\
\mb{b}\left(t_k\right)^\mr{T} & 0
\end{bmatrix}}\right)},
\end{equation}
where $t_k=(k-1)\delta\tau$.

We decompose this propagator as
\begin{equation}
\tilde{U}(\tau)=
\begin{bmatrix}
U\left(\tau\right) & \mb{0} \\
\mb{U}_b^{\mr{T}}\left(\tau\right) & 0
\end{bmatrix}.
\end{equation}
Then, we impose the boundary conditions in Eq.~\eqref{eq:solution} to get
\begin{equation}
\mb{p}\left(\tau\right)=\mb{p}\left(0\right)=U\left(\tau\right)\mb{p}\left(0\right).
\end{equation}
The calculation of $\mb{p}$ has been reduced to determining the eigenvector of $U\left(\tau\right)$ corresponding to the eigenvalue $1$.
Finally, we calculate the average charge according to Eq.~\eqref{eq:solution} as $\left\langle q\right\rangle\left(\tau\right)=\mb{U}_b\left(\tau\right)\cdot\mb{p}(0)$.
As a result, the average current can be written as $I=\left\langle q\right\rangle\left(\tau\right)/\tau$.

\end{document}